\documentclass[apj,numberedappendix]{emulateapj}

\usepackage{xcolor}
\usepackage{amsmath}
\usepackage{natbib}
\usepackage{graphicx}
\usepackage{epstopdf}
\usepackage[colorlinks=true,citecolor=blue,linkcolor=blue,urlcolor=blue]{hyperref}

\newcommand{\msunh}{\>h^{-1}\rm M_\odot}

\newcommand{\Msun}{\>{\rm M_{\odot}}}

\newcommand{\mpch}{\>h^{-1}{\rm {Mpc}}}

\newcommand{\msunhh}{\>h^{-2}\rm M_\odot}
\newcommand{\kpch}{\>h^{-1}{\rm {kpc}}}

\usepackage{color}
\definecolor{darkgreen}{rgb}{0.0,0.5,0.0}
\usepackage{array}
\usepackage{booktabs}
\usepackage{multirow}

\shorttitle{EG test with SDSS}
\shortauthors{Luo et al.}

\begin{document}


\title{Emergent Gravity fails to explain color-dependent galaxy-galaxy lensing signal from SDSS DR7}

\author{Wentao Luo\altaffilmark{1,2}, Jiajun Zhang\altaffilmark{3}, Vitali Halenka\altaffilmark{4}, Xiaohu Yang\altaffilmark{5,6}, Surhud More\altaffilmark{7,1}, Chris Miller\altaffilmark{3,8}, Lei Liu\altaffilmark{9}, Feng Shi\altaffilmark{10}}
 
\altaffiltext{1} {Kavli Institute for the Physics and Mathematics of the Universe (Kavli IPMU, WPI), University of Tokyo, Chiba 277-8582, Japan; wentao.luo@ipmu.jp}
\altaffiltext{2}{CAS Key Laboratory for Research in Galaxies and Cosmology, University of Science and Technology of China, Hefei, Anhui 230026, China}
\altaffiltext{3} {Center for Theoretical Physics of the Universe, Institute for Basic Science(IBS), Daejeon 34126, Korea}

\altaffiltext{4}{Department of Physics, University of Michigan, Ann Arbor, MI 48109 USA}

\altaffiltext{5} {Department of Astronomy, School of Physics and Astronomy, and Shanghai Key Laboratory for Particle Physics and Cosmology,  Shanghai Jiao Tong University, Shanghai, 200240, China}
  
\altaffiltext{6} {Tsung-Dao Lee Institute, and Key Laboratory for
    Particle Physics, Astrophysics and Cosmology, Ministry of Education, Shanghai Jiao Tong University,
  Shanghai, 200240, China}
  
\altaffiltext{7} {The Inter-University Center for Astronomy and Astrophysics, Post bag 4, Ganeshkhind,
Pune, 411007, India}
  
\altaffiltext{8}{Department of Astronomy, University of Michigan, Ann Arbor, MI 48109, USA}

\altaffiltext{9} {Shanghai Astronomical Observatory, Shanghai 200030, China}

\altaffiltext{10}{Korea Astronomy and Space Science Institute
776 Daedeok-daero, Yuseong-gu, Daejeon 34055, Republic of Korea}

\begin{abstract}
We test Verlinde's Emergent Gravity (EG) theory using galaxy-galaxy lensing technique based on SDSS DR7 data. In the EG scenario, we do not expect color dependence of the galaxy sample in the `apparent dark matter' predicted by EG, which is exerted only by the baryonic mass. If the baryonic mass is similar, then the predicted lensing profiles from the baryonic mass should be similar according to EG, regardless of the color of the galaxy sample.  We use the stellar mass of the galaxy as a proxy of its baryonic mass. We divide our galaxy sample into 5 stellar mass bins, and further classify them as red and blue subsamples in each stellar mass bin. If we set halo mass and concentration as free parameters, $\Lambda$CDM is favored by our data in terms of the reduced $\chi^2$,  while EG fails to explain the color dependence of ESDs from galaxy-galaxy lensing measurement.
\end{abstract}



\section{Introduction}
\label{sec_intro}
Today, the concordance cosmological model where dark matter and dark energy form about 95 percent of the energy density of the Universe is supported by a plethora of observations including those of the Cosmic Microwave Background (CMB) \citep[see e.g.,][]{planck2016}, Supernovae of Type Ia \citep[see e.g.,][]{Perlmutter1999}, Baryon Acoustic Oscillations (BAO) \citep[see e.g.,][]{Eisenstein2005} as well as weak lensing \citep[see e.g.,][]{CFHTlens,KiDS,Shi2017}. The observational data from the above probes can be described by merely half a dozen major parameters, a.k.a $\Lambda$CDM, despite a recent claim of 5.3$\sigma$ tension in $H_0$ between CMB probe \citep{planck2018} and strong lensing time delay project H0LiCOW \citep{wong2019},  SH0ES project \citep{riess2016ApJ}. Regardless of this success, the dark matter still remains a 
mystery.

The concept of dark matter was first introduced by \cite{Zwicky1937} based on the anomalous dynamics of galaxies in clusters, which required excess gravitational influence than that from the baryonic component only. Observations of galaxy rotation curves \citep{Bosma1981,Rubin2001} further confirm this anomalous behaviour. These observations require the presence of dark matter that can not be detected in any electromagnetic observations which dominates the matter sector of the Universe. Since then, the study of the properties of dark matter has become one of the frontier fields from both particle physics perspective and modified gravity scenario.

There are plenty of models from particle physics and possible detection experiments in literature ranging from light boson model [e.g. axion dark matter \citep{duffy2009NJPh}, which arises from the Peccei-Quinn solution] to the strong CP problem [sterile neutrino as potential candidate \citep{kisslinger2019IJMPA}] and weakly interacting massive particles  predicted by R-parity-conserving supersymmetry \citep{jungman1996PhR}. And so far, there are no experiments that can confirm any of the models, neither earth based labs \citep{kang2010,zhang2019SCPMA,xenonPhysRevLett} nor space based detection \citep{fermin2020PhRvD,dampe2019RAA}

On the other hand, some researchers try to view dark matter as the modification of the theory of gravity. For example, MOdified Newtonian Dynamics or MOND \citep{milgrom1983,milgrom2011arXiv,milgrom2020SHPMP} explains the high speed stars in galaxies by adding interpolation function to modify the acceleration of Newtonian theory. \cite{bekenstein2004PhRvD} further improves MOND by considering gravity as a mixture of dynamics of metric, a scalar, and a 4-vector field, a.k.a TeVes, which can predict consistent weak lensing signals. \cite{milgrom2013PhRvL} claims that MOND prediction agrees with the velocity dispersion to r band luminosity relation $\sigma-L_r(h^{-2}L_{\odot})$ based on the CFHT data \citep{heymans2013MNRAS}, but without comparison of the galaxy-galaxy lensing profiles directly as in \cite{brouwer2017}. \cite{chae2020ApJ} finds evidence that supports MOND gravity from the observations of Spitzer Photometry and Accurate Rotation Curves (SPARC).

Among the various MOND models, there is a unique one based on an entropic scenario. \cite{verlinde2016} reconsidered the gravity as the underlying microscopic description inspired by the laws of black hole thermodynamics \citep{bardeen1973}, i.e. Emergent Gravity (EG). \cite{brouwer2017} firstly tested this assumption using galaxy-galaxy lensing technique based on the data from KiDs \citep{derong2013} and GAMA \citep{gama2009}, they claimed that both dark matter scenario and EG can fit the galaxy-galaxy lensing signal equally well.

\cite{zuhone2019} tested Emergent Gravity using relaxed galaxy clusters and found that inclusion of the central galaxy improves agreement between observations and the theory in the inner regions $(r \leq 30)\,{\rm kpc}$. On larger scales, the predictions are discrepant with observations and $\Lambda$CDM models fit the observations better. However, \cite{halenka2018} found that there is enough freedom in the EG theory for it to agree with the data as well as $\Lambda$CDM, especially after accounting for possible observational systematics. Baryonic physics complicates the inference of the underlying gas density profile and weakens the constraining power of observations.

In this paper, we re-test this theory by using a much larger survey data from Sloan Digital Sky Survey(SDSS) DR7 \citep{abazajian2009} as well as two cosmology models in $\Lambda$CDM framework, i.e. WMAP5 \citep{komatsu2009} and PLANCK18 \citep{planck2018}. We minimize the complicated modeling of massive clusters by only selecting single galaxy systems from the \cite{yang2007} catalog with mean halo mass  $\log{M} \leq 13.5 h^{-1} M_\odot$. None of the systems have X ray detection, which further minimizes the hot baryonic contribution. With this data set, we are able to select isolated galaxies. Our sample is at least 5 times bigger than that used in \citep{brouwer2017} as we use the group catalog built by \citep{yang2007}. The models of galaxy-galaxy lensing signals from both EG and $\Lambda$CDM are described in Sec. \ref{models}. We introduce the lensing data and methodology in Sec. \ref{signals}. The results are given in Sec. \ref{results}. Finally, we summarize and discuss in Sec. \ref{summary}.


\section{the galaxy-galaxy lensing models}
\label{models}

\subsection{Lensing model in Emergent Gravity}
\label{modeleg}

The tangential distortions of background galaxy shapes caused by weak gravitational lensing are proportional to the excess surface density (ESD), $\Delta\Sigma$, which is the difference in the average surface density within a projected radius R and the surface density at radius R. The ESD is related to the tangential shear $\gamma_t(R)$
by a factor $\Sigma_c$
\begin{equation}
\label{eq:esd}
\gamma_t(R)\Sigma_c=\Delta\Sigma(R)=\Sigma(\leqslant R)-\Sigma(R),
\end{equation}
where $\Sigma_c$ is the critical density dependent upon the geometric distances between the observer, lens and the source galaxy. For the $\Lambda$CDM case, we refer to \cite{yang2006} for detailed formulation, which is well established in galaxy-galaxy lensing studies. 

In Emergent Gravity (hereafter EG) scenario, a term additional to the normal baryonic mass arises and that can act as an apparent dark matter contribution. Based on \cite{verlinde2016}, the extra term of gravitational potential is exerted by the entropy displacement from total galaxy mass $M_g(r)$, where $M(r)$ is the mass enclosed within a radius $r$. This mass includes stellar mass and cold gas components. As a result, the apparent mass $M_a(r)$ is related to $M_g(r)$ via 
\begin{equation}
\label{eq:egmass}
M_a^2(r)=\frac{cH_0r^2}{6G}\frac{d[M_g(r)r]}{dr}.
\end{equation}
As  in \cite{brouwer2017}, for a typical mass of $M=10^{10}h^{-2}M_{\odot}$, EG becomes significant over scale larger than $2\kpch$. We measure our galaxy-galaxy lensing signal from $0.01\mpch$ all the way to $1\mpch$ to empirically test the scale dependence of both theories. We follow \cite{brouwer2017} that beyond $30\kpch$, the galaxy can be considered a point mass. We exclude the first data point within $30 \kpch$. In Sec. ~\ref{results}, we calculate the $\chi^2$ excluding the first data point of each of the measurements below this scale.


From Eq. \ref{eq:egmass}, we get the mass distribution
\begin{equation*}
    M_a(r) = \Bigr[ \frac{cH_0 r^2}{6G} \Bigr(M_g(r) + r\frac{\partial M_g(r)}{\partial r} \Bigr) \Bigr]^{0.5}
\end{equation*}
and the second term on the right hand side is gone under the point mass assumption, i.e. $M_g(r)=M_g$
and we can treat the factor $\sqrt{\frac{cH_0}{6G}}$ as a combined constant $C_a$, also following \cite{brouwer2017}. The density profile can be related to the derivative of the mass distribution
\begin{equation}
\rho_{EG}(r)=\frac{1}{4\pi r^2}\frac{dM_a(r)}{dr}=\frac{C_a\sqrt{M_g}}{4\pi r^2}.
\end{equation}
The 2D surface density at projected distance R is then bearing the form
\begin{equation}
\label{eq:pmass}
\Sigma_{EG}(R)=\int_{-\infty}^{\infty}\rho_{EG}(R,\chi)d\chi=\frac{C_a\sqrt{M_g}}{4R},
\end{equation}
where $r^2=R^2+\chi^2$ with $R$ as the projected distance and $\chi$ as the distance along the line of sight. Then the ESD of EG point mass can be calculated
\begin{equation}
\Delta\Sigma_{EG}(R)=\frac{C_a\sqrt{M_g}}{4R},
\end{equation}
which happens to be the same as Eq. \ref{eq:pmass}. Together with the original baryonic mass contribution, the total ESD profile as predicted by EG is
\begin{equation}
\Delta\Sigma(R)_{all}=\frac{M_g}{\pi R^2}+\Delta\Sigma_{EG}(R).
\end{equation}

In the $\Lambda$CDM scenario, the dark matter density profile can be accurately described by an NFW profile \citep{nfw1997}. When converting the 3D NFW profile to the 2D ESD, it differs from the EG profile. 


\subsection{Lensing model in $\Lambda$CDM}

We model the the ESD based on the NFW density profile with two free parameters namely, halo mass and concentration parameters, and we label this model as 'NFW'. 
We use \cite{yang2006} formulation to model the ESD given a halo mass based on
an NFW dark matter halo profile \cite{nfw1997}, 
\begin{equation}
   \rho(r)=\frac{\rho_0}{(r/r_s)(1+r/r_s)^2},
\end{equation}
with $\rho_0=\frac{{\bar\rho\Delta_{vir}}}{3I}$, where
$\Delta_{vir}=200$, $I=\frac{1}{c^3}\int_0^c
\frac{xdx}{(1+x)^2}$. Here $c$ is the concentration parameter defined
as the ratio between the virial radius of a halo and its
characteristic scale radius $r_s$.

Recently, the group catalog was also updated to include abundance matching based halo mass estimation in both the WMAP5 and PLANCK18 cosmology. We will therefore further examine the cosmology dependence.

In $\Lambda$CDM scenario, the ESD is composed of the following simple two components: host
halo mass and the
stellar mass,
\begin{equation}
  \Delta\Sigma(R)=\Delta\Sigma_{host}(R)+\Delta\Sigma_{*}.
\end{equation}
We do not include two halo term, which is the signal caused by the large scale structure due to the fact that
we select the isolated galaxies and we only measure our signal to $1\mpch$. 
The contribution of the stellar components from the lens galaxy can be modeled as a point mass
\begin{equation}
\label{eq:stellar}
    \Delta\Sigma_{*}(R)=\frac{M_{*}}{\pi R^2}.
\end{equation}

$\Delta\Sigma_{host}$ is the contribution of the halo given that the galaxy is perfectly located at the center. 




\section{The Galaxy-Galaxy Lensing Signals}
\label{signals}

In this section, we describe the data we use to measure the galaxy-galaxy lensing
signals. 

\subsection{Lenses}

\begin{figure}
\includegraphics[width=0.48\textwidth]{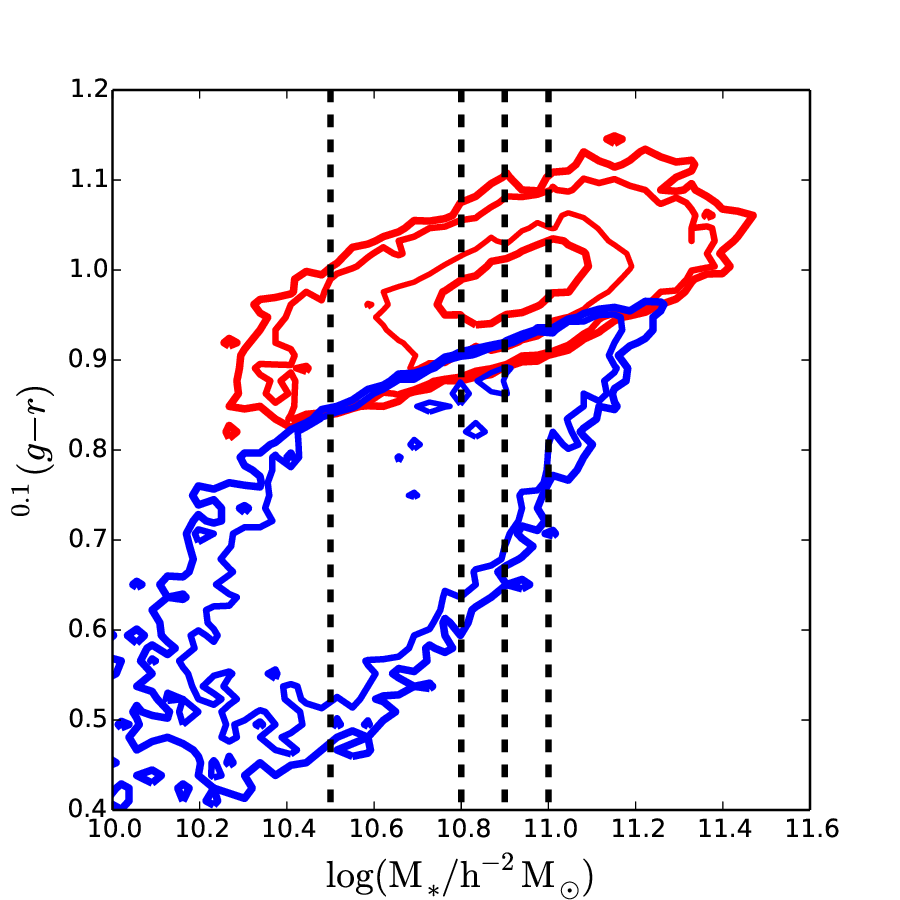}
\caption{This is the 2D distribution contour plot between the color and stellar mass of lens galaxy sample. The dashed vertical lines divide the plot into 5 stellar mass regions, each region is further divided into red and blue sub samples. The overlap region between the blue and red are due to the fact that the threshold is calculated using color and r band magnitude rather than stellar mass. }\label{fig:bins}
\end{figure}

The lenses are selected from the galaxy group catalog constructed from the spectroscopic SDSS survey (DR7) \citep{yang2007}, 
which is based on a halo-based group finding algorithm \citep{yang2005MNRAS}. Recently, \cite{yang2020arXiv} extended this  group finder so that it can deal with galaxies with photometic and spectroscopic redshifts simultaneously, and successfully applied it to the DESI Legacy surveys data release 8 \citep{desi2019AJ}. The strength of this algorithm is related with its iteration nature and the application of an adaptive filter according to the general properties of the dark matter halos. It starts with assuming each galaxy is a potential group candidate and then calculates the total luminosity of each system. The halo mass is then estimated based on abundance matching. After the halo mass estimation, the other quantities such as velocity dispersion, and virial radius e.t.c are then deduced. The member galaxies are determined by selecting galaxies that meet the criteria, which include distance and redshift information. All the above procedures are iterated several times till the mass-to-light ratios converge. There are systems with only one central galaxy, meaning that there are no other galaxies brighter than the  magnitude limit $r=17.77$ within projected virial radius and with $\Delta z=|z_i-z_{group}|$ less than the viral velocity of the dark matter halo along the line-of-sight direction.

In total, there are 472419 groups in the sample. In order to minimize the effects of nearby structures, we only select single galaxy systems which further reduce the number to 400608. The stellar mass of each galaxy is computed using stellar mass-to-light ratio and color from \cite{bell2003}, but with a Kroupa IMF \citep{kroupa2001}.  This leads to a -0.1 correction to the stellar mass-to-light ratio relation. The statistical scatter of color-based stellar M/L ratio is about 20\%. Systematics rising from galaxy age, dust and bursts of star formation in total contribute $\sim$0.1dex scatter. 
In general, the scatter may induce some Eddington bias to the average stellar mass of galaxies. However, since the total amount of scatter is quite small, the overall Eddington bias can only leads to $\sim$0.03dex overestimation of stellar mass,  which  will not impact any of our results significantly.

The sample is sub-divided into different stellar mass bins following \cite{brouwer2017}. We add one more stellar mass bin compared to their study, with $\log M_{st}$ mass $\geq$ 11.0 due to the larger sample size. The mean redshift of our sample is lower than \cite{brouwer2017}, so our work is complementary to theirs as low z test and it provides better agreement with a small redshift assumption of EG model. Moreover, our samples are at least five times larger to improve the measurement. 

The vertical dashed lines in Fig. 1 divide our sample in 5 $M_{*}$ bins. We further sub-divide our sample of galaxies in to blue star forming galaxies and red passive galaxies based on a cut in the color magnitude plane from \cite{yang2008} such that
\begin{equation}
\label{eq:rb}
 \mathrm{  ^{0.1}(g-r)=1.022 - 0.0652 x - 0.0031 x^2 }, 
\end{equation}
where $x = {\rm ^{0.1}M_r} - 5 \log h + 23.0$, and ${\rm ^{0.1}M_r} - 5 \log h$ is the absolute magnitude of galaxy after $K$ correction and evolution correction to redshift z=0.1.  The statistics of the our sub-samples is given in Table.~\ref{tab:tbl-1} and illustrated in Fig. 1. The overlap between the red and blue contours are due to 
the fact that threshold in Equation~\ref{eq:rb} is calculated based on color and
magnitude, while Fig.~\ref{fig:bins} is the color and stellar mass 2D distribution.

We treat the gas contribution following \cite{brouwer2017, boselli2014} for the blue galaxies, which applies a factor $\rm f_{cold}$ so that the total galaxy mass $M_g$ can be written as
\begin{equation}
  \rm  M_g=M_*(1+f_{cold}).
\end{equation}
\cite{boselli2014} gives an empirical form of $\rm f_{cold}$ based on Herschel Reference Survey \citep{boselli2010}
\begin{equation}
   \rm log(f_{cold})=-0.69log(M_*/h^{-2}M_{\odot})+6.63.
\end{equation}
For the red galaxy, we apply a constant fraction of 1\%, which is the 
upper limit from \cite{boselli2014} for early-type galaxies. 
We do not take the hot gas into consideration so far because firstly, the dominant factor is stellar mass as in \cite{brouwer2017}, as we focus on the point mass contribution by selecting single galaxy system, and the hot gas contribution is less than the 0.1dex systematic for the stellar mass estimation.

We also add the fitted NFW halo mass for each sample with errors in Sec.~\ref{results}. 

\begin{table*}
\label{tab:1}
\begin{center}
  \caption{\label{tab:tbl-1} Properties of the lens samples
    created for this paper. $\mathrm{log(Mh_{W5}/\msunh)}$ and $\mathrm{log(Mh_{P18}/\msunh)}$ are the weak lensing fitted mass for the two different cosmologies.}
\begin{tabular}{llcccc}
  \hline \\  
  $\mathrm{\log M_{st}}$ range & Num & $\mathrm{\langle z \rangle} $ & $\mathrm{\langle 
\log (M_{st}/\msunhh) \rangle}$ & $\mathrm{log(Mh_{W5}/h^{-1}\Msun)}$ & $\mathrm{log(Mh_{PL}/h^{-1}\Msun)}$ \\
  \hline \\
  8.5-10.5    & 216 212 & 0.078 & 10.001  & $11.563^{+0.059}_{-0.062}$ & $11.686^{+0.063}_{-0.069}$\\
  RED         & 69 914  & 0.074 & 10.180  & $11.861^{+0.067}_{-0.073}$ & $11.983^{+0.070}_{-0.076}$ \\
  BLUE        & 146 298 & 0.079 & 9.916   & $11.354^{+0.099}_{-0.112}$ & $11.378^{+0.099}_{-0.113}$ \\
  \hline
  10.5-10.8   & 104 484 & 0.123 & 10.648  & $11.935^{+0.085}_{-0.087}$ & $12.210^{+0.072}_{-0.077}$\\
  RED         & 61 278  & 0.115 & 10.654  & $12.086^{+0.108}_{-0.108}$ & $12.284^{+0.089}_{-0.093}$ \\
  BLUE        & 43 206  & 0.134 & 10.640  & $11.761^{+0.149}_{-0.187}$ & $11.758^{+0.161}_{-0.207}$ \\
  \hline
  10.8-10.9   & 28 747  & 0.143 & 10.848  & $12.493^{+0.121}_{-0.119}$ & $12.725^{+0.103}_{-0.105}$\\
  RED         & 19 735  & 0.140 & 10.849  & $12.566^{+0.108}_{-0.108}$ & $12.810^{+0.104}_{-0.107}$ \\
  BLUE        & 9 012   & 0.151 & 10.847  & $12.346^{+0.367}_{-0.546}$ & $12.312^{+0.399}_{-0.585}$ \\
  \hline
  10.9-11.0   & 22 330  & 0.155 & 10.946  & $12.449^{+0.189}_{-0.225}$ & $12.596^{+0.220}_{-0.247}$ \\
  RED         & 16 965  & 0.155 & 10.948  & $12.516^{+0.187}_{-0.218}$ & $12.948^{+0.465}_{-0.569}$ \\
  BLUE        & 5 365   & 0.156 & 10.944  & $12.218^{+0.506}_{-0.690}$ & $12.601^{+0.228}_{-0.271}$ \\
  \hline
  11.0-above  & 24 717  & 0.165 & 11.087  & $12.673^{+0.104}_{-0.102}$ & $13.000^{+0.075}_{-0.083}$\\
  RED         & 20 584  & 0.166 & 11.119  & $12.733^{+0.106}_{-0.103}$ & $13.075^{+0.081}_{-0.086}$ \\
  BLUE        & 4 133   & 0.158 & 11.096  & $12.155^{+0.411}_{-0.578}$ & $12.426^{+0.384}_{-0.656}$ \\
  \hline
\end{tabular}
\end{center}
\end{table*}

\subsection{Sources and estimator}

For the source catalog, we use the shape catalog created by \cite{luo_gg1} based on SDSS DR7 imaging data. The DR7 imaging data, with \textit{u}, \textit{g}, \textit{r}, \textit{i} and \textit{z} band, covers about 8423 square
degrees of the LEGACY sky ($\sim$230 million distinct photometric objects). The total number of objects identified as galaxies is around
150 million.  
The final shape catalog for our study contains about 40 million galaxies with position, shape, shape error and photoZ information based on \cite{csabai2007}, which fits a local color-color hyper-plane with nearest 100 objects.

The shear signals $\Delta\Sigma(R)$  can be
measured by the weighted mean of source galaxy shapes,
\begin{equation}
\mathrm{\Delta\Sigma(R)=\frac{1}{2\bar{R}}\frac{\sum w_ie_t(R)\Sigma_{cls}}{\sum w_i}},
\end{equation}
where $w_i$ is the weight for each source galaxy. 
$\Sigma_{cls}=\frac{c^2}{4\pi G}\frac{D_s}{D_{ls}D_l(1+z_l)^2}$ is 
the critical density for each lens-source pair. We measure the signal in 6 
equal logarithm bins in projected co-moving distance from 0.01Mpc/h to 1Mpc/h.
The weighting term is
composed by shape noise $\sigma_{\rm shape}$ and that from sky
$\sigma_{\rm sky}$
\begin{equation}
\label{eq:weig}
w=\frac{1}{(\sigma_{\rm sky}^2+\sigma_{\rm shape}^2)\Sigma_{cls}^2}.
\end{equation}

We correct the dilution effect by calculating the boost factor, which is from the  contamination of non-lensed galaxies due to inaccurate photometric redshift 
\begin{equation}
   \mathrm{ B(R)=\frac{N_{rand}}{N_{lens}}\frac{\sum_{i}^{N_{lens}}w_{ls}}{\sum_{j}^{N_{rand}}w_{rs}}}.
\end{equation}
$N_{lens}$ and $N_{rand}$ are the number of lens galaxy of each sample and corresponding
random sample. The weights $w_{ls}$($w_{rs}$) correspond to each lens (random position, N(zrand)=N(zlens)) as in Eq. \ref{eq:weig}.

The $\chi^2$ can be calculated as
\begin{equation}
    \chi^2=\mathrm{((data-model)^TC^{-1}(data-model))},
\end{equation}
where $C^{-1}$ is the inverse covariance matrix. We further add photometric redshift systematic from weak lensing measurement to the trace of covariance matrix when we calculate the $\chi^2$. We estimated the systematics caused by photometric redshift to be 2.7\% \citep{luo_gg1} for the most massive stellar mass bin.

\section{Results}
\label{results}
In this section, we describe the results from the comparison between the EG and $\Lambda$CDM model. Our use of a larger data set, allows us to obtain high SNR measurement of galaxy-galaxy lensing signals even after we split the sample into red and blue lens samples to study the color dependence. The SNR is ranging from 17.6 for blue galaxy sample to 28.1 for red galaxy sample based on EQ.(5) in  \cite{leauthaud2017}.

Fig. 2 is the comparison between the data
and different models, i.e. NFW (Mh, c as free parameters) and Emergent Gravity (EG). It is well known that the lensing signal is dependent on several cosmological parameters, e.g. $\Omega_m, \sigma_8$ and Hubble parameter. Whereas EG depends only on Hubble parameter as shown in Eq.~\ref{eq:egmass}. That is why EG shows stronger cosmology variance than $\Lambda$CDM in terms of reduced $\chi^2$. Apparently, EG prefers PLANCK18 cosmology with reduced $\chi^2=1.907$ to WMAP5(reduced $\chi^2=2.959$) as in Table. ~\ref{tab:tbl2}.  We exclude the first data points from all measurements because it is below 30kpc/h, but still
show the $\chi^2$ in table ~\ref{tab:tbl2} (inside the parenthesis) by including the first data points to see the difference.

Our measurement at small stellar mass bins have very high signal to noise ratio. And
due to the selection of isolated systems, we have less contribution from adjacent 
structure. Therefore, the decreasing feature in the first two stellar mass bins play
an important role to the whole $\chi^2$. We do not use the extended model as in
\cite{brouwer2017}, because the extended model only make the $\chi^2$ larger.

\begin{figure*}
\includegraphics[width=0.5\textwidth]{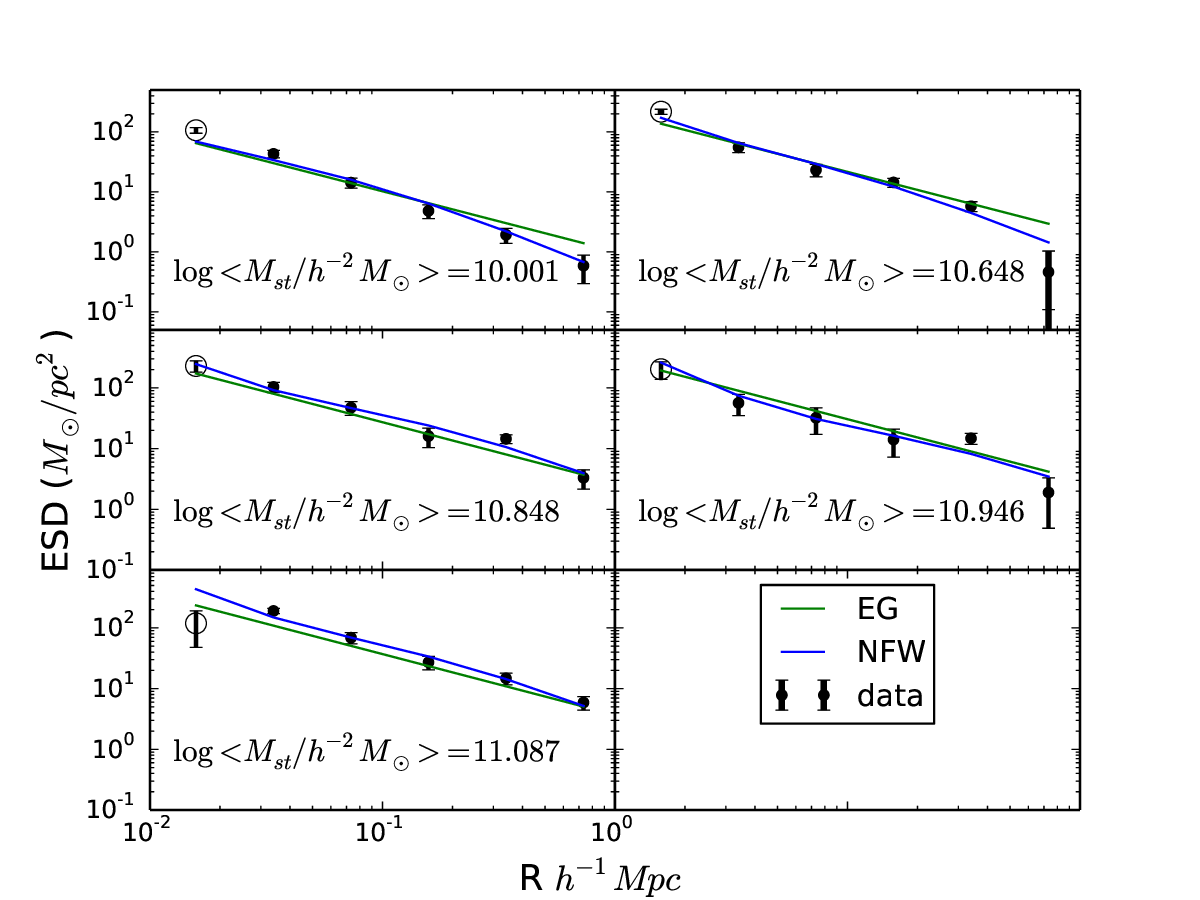}
\includegraphics[width=0.5\textwidth]{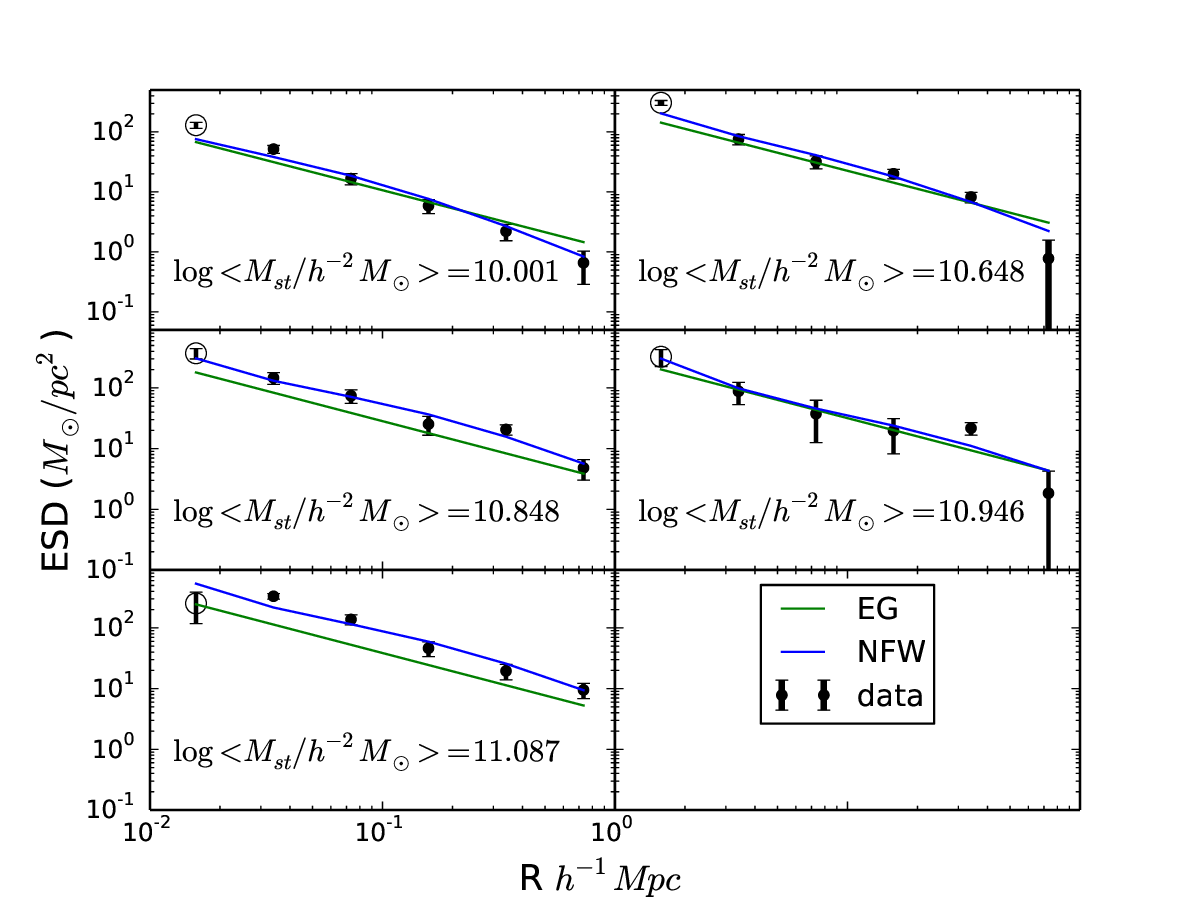}
\caption{\textit{Left}: The prediction of emergent gravity is shown in green and the prediction of $\Lambda$CDM model is shown in  blue with PLANCK18. Comparing to the weak lensing signal shown in black dots with errorbars. \textit{Right}: Same as left figure but with WMAP5 cosmology. The excluded data points in our analysis are shown as empty circles at scale smaller than 30$\kpch$}\label{fig:signal}
\end{figure*}

\begin{figure*}
\centering
\includegraphics[width=5cm,height=5cm]{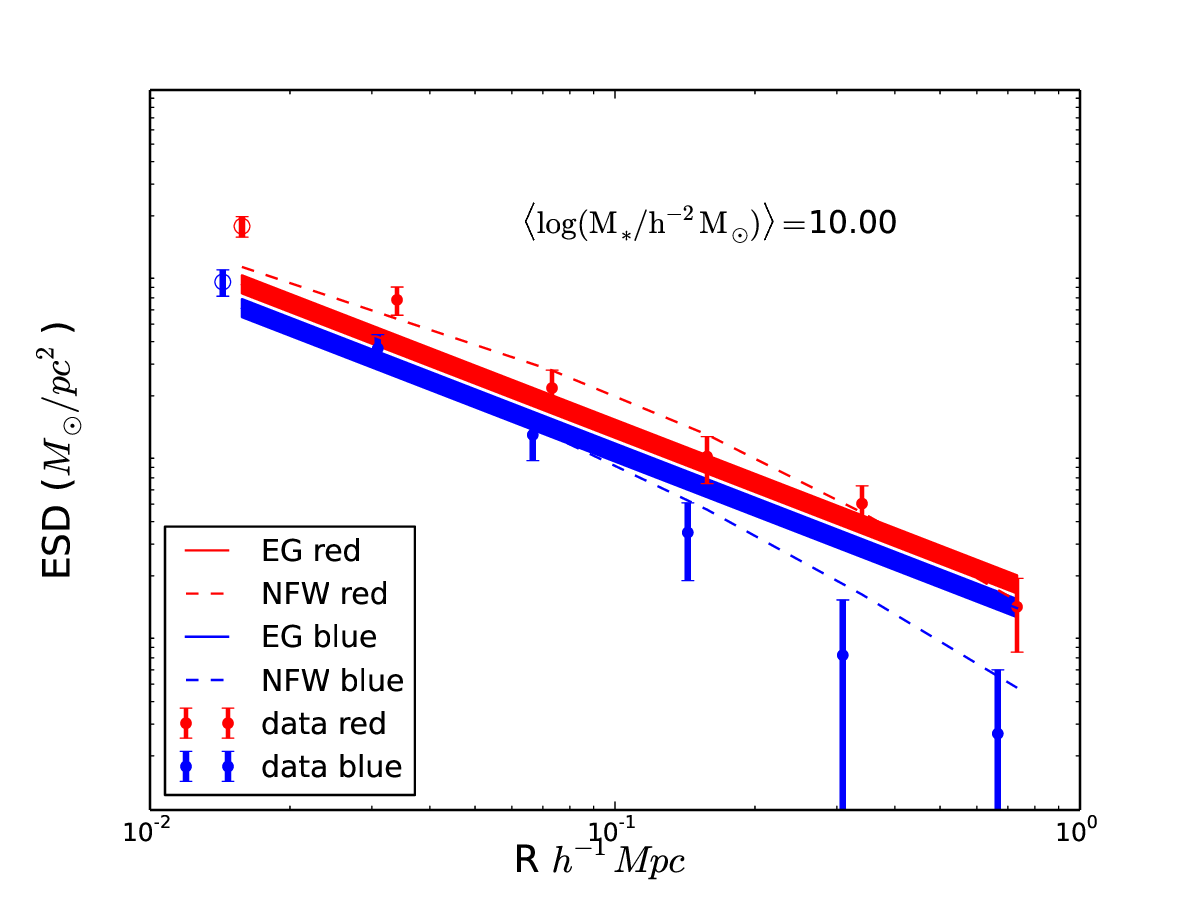}
\includegraphics[width=5cm,height=5cm]{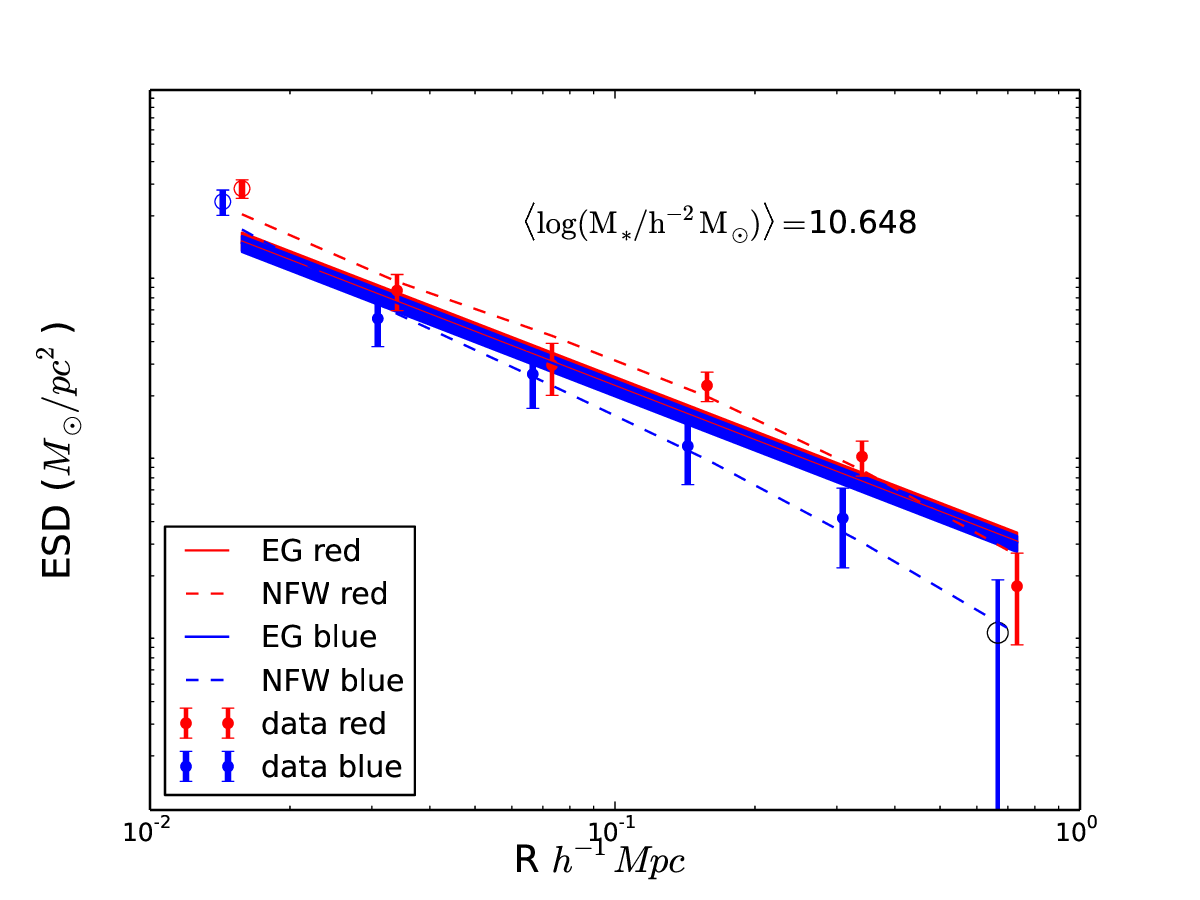}
\includegraphics[width=5cm,height=5cm]{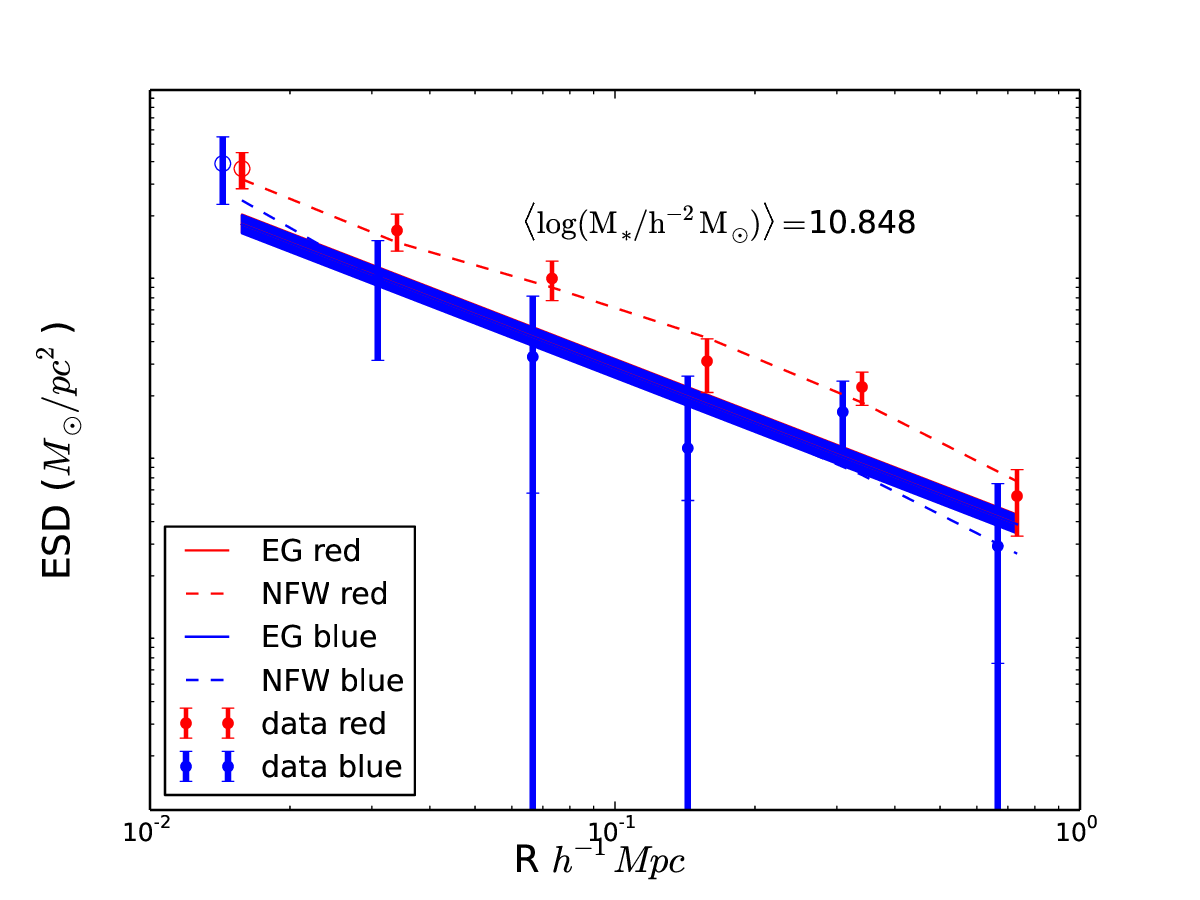}
\caption{From left to right, these are the plots of stellar mass bin 1, 2 and 3 based on 
PLANCK18 cosmology. The red and blue dots are the measurement from red and blue galaxy samples. We exclude the first data points within 30$\kpch$ as empty circles. The empty circle at large scale in the middle panel denotes a negative value. The red and blue
solid lines are the EG model, dashed lines are from NFW model. The bandwidth from EG model is due to the 0.1dex systematic from stellar mass estimation based on the method of \cite{bell2003}. }\label{fig:finalC}
\end{figure*}

\begin{figure}
\includegraphics[width=0.5\textwidth]{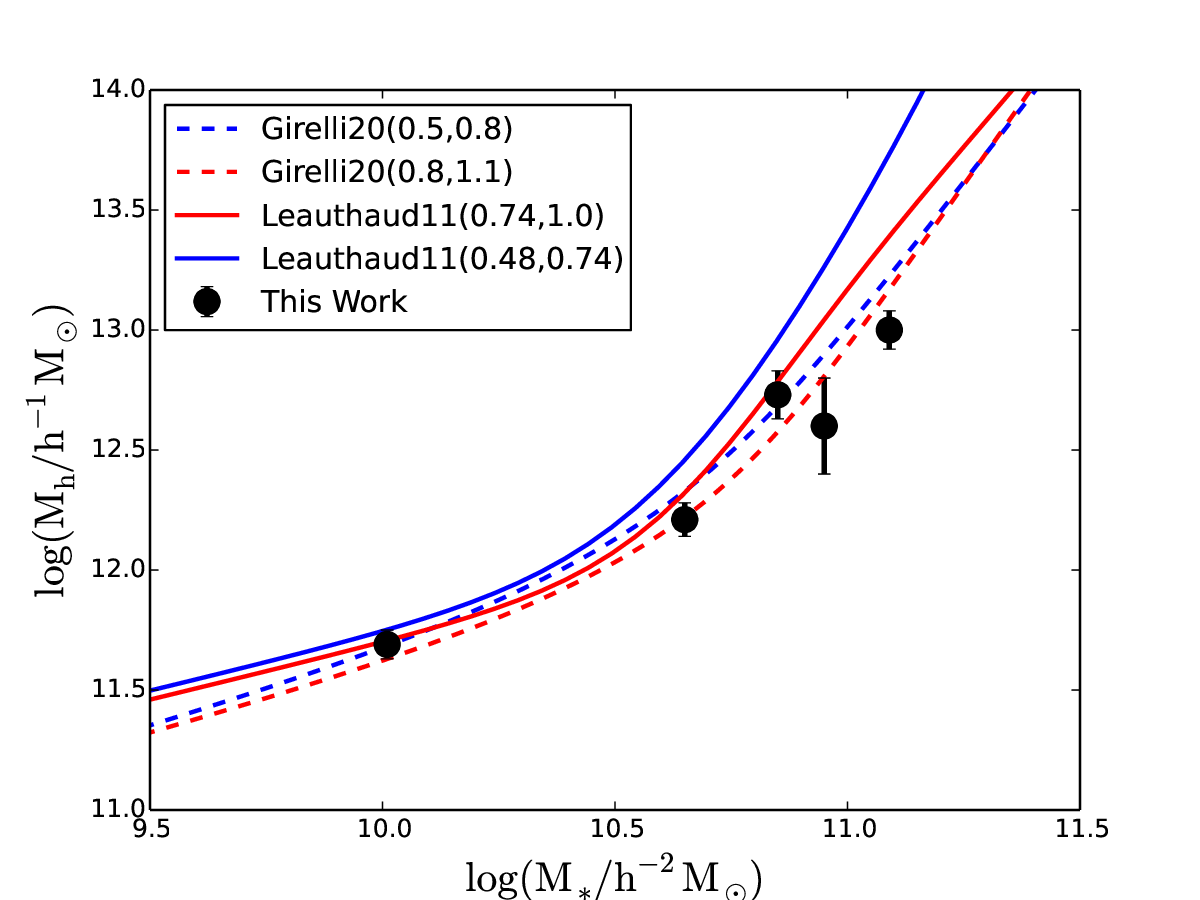}
\caption{Stellar Mass to Halo Mass Relation (SHMR) in our study. Notice that we select single system groups. So at massive end, it is not consistent with the theoretical curve due to this selection. And the modeling is simply based on an NFW and stellar mass with fixed stellar mass contribution.}\label{fig:msmh}
\end{figure}

We show the color dependence in PLANCK18 cosmolgy 
in Fig. \ref{fig:finalC}. 
The NFW model with free halo mass and concentration,
apparently is favored by the data, especially the blue data.
Fig. \ref{fig:finalC} shows the ESD profile from the first three stellar mass bins in PLANCK18 cosmology. 
Due to larger signal to noise ratio, the rest of two ESD profiles from massive stellar mass bins do not carry so much information. 

In the left panel of Fig. \ref{fig:finalC}, there is significant difference between
the ESDs from red and blue lenses. The ESD from the red lens is larger than their
blue counterpart with 0.164dex difference in stellar mass but 0.605dex difference in halo mass in PLANCK18 cosmology. 
The stellar mass difference shrinks to 0.014dex, but the halo mass difference is 0.526dex for the second 
stellar mass bin sample. The third stellar mass bin sample has almost identical stellar mass for blue and
red galaxy, but the halo mass difference is still up to 0.498dex.

Comparing to the halo masses directly provided in the group catalog, the first three stellar mass bins have consistent halo mass estimation
for the whole sample after considering 0.07 Eddington bias estimated
from \cite{luo2018ApJ}. The last two shows significant discrepancy with
abundance matching halo mass, ~0.5dex difference in the last stellar mass bin.
We attribute this to the selection effect
that we only select single galaxy system. Fig.~\ref{fig:msmh} shows the Stellar mass to Halo Mass Relation (SHMR) of our measurement. Our measurement agrees well with both observational calibration \citep{leauthaud2017} and simulation calibration \citep{girelli2020A&A} except for the most massive stellar mass bin. That is due to our simple NFW model and the selection of single galaxy systems. The multi-galaxy systems
in stellar mass bin 4 is about 20.3\% and 33.4\% for stellar mass
bin 5. We re-calculate the multi-galaxy sample halo mass for
those two bins in PLANCK18 cosmology and obtain higher halo mass
than the single systems in the same stellar mass bin, which are
12.873 and 13.533 respectively. If we simply take the weighted
average halo mass together with single systems, we get 12.654$\rm ~\pm 0.23dex$ and 13.178$\rm ~\pm 0.08dex$, vs 12.985 and 13.299 from abundance matching.


\begin{table}[h!]
\begin{center}
  \caption{\label{tab:tbl2} $\chi^2$ comparison between EG and $\Lambda$CDM. The $\chi^2$ values in the parenthesis are calculated by including the first
  data points from the measurements.} 
\begin{tabular}{lcc}
  \hline
\\
  Cosmology & NFW ($\chi^2/dof_{=15}$) &  EG($\chi^2/dof_{=25}$)\\
\\  
  \hline  
  \\
 WMAP5    & 0.949(1.453)   & 2.959(3.739) \\
 RED      & 0.717(1.433)   & 1.851(3.397) \\
 BLUE     & 0.731(0.682)   & 2.441(2.085) \\
 \hline
 PLANCK18 & 0.868(0.966)   & 1.907(1.770)\\
 RED      & 0.718(0.885)   & 1.792(1.762) \\
 BLUE     & 0.659(0.626)   & 2.730(2.391) \\
  \hline
\end{tabular}
\end{center}
\end{table}

\begin{figure}
\includegraphics[width=0.5\textwidth]{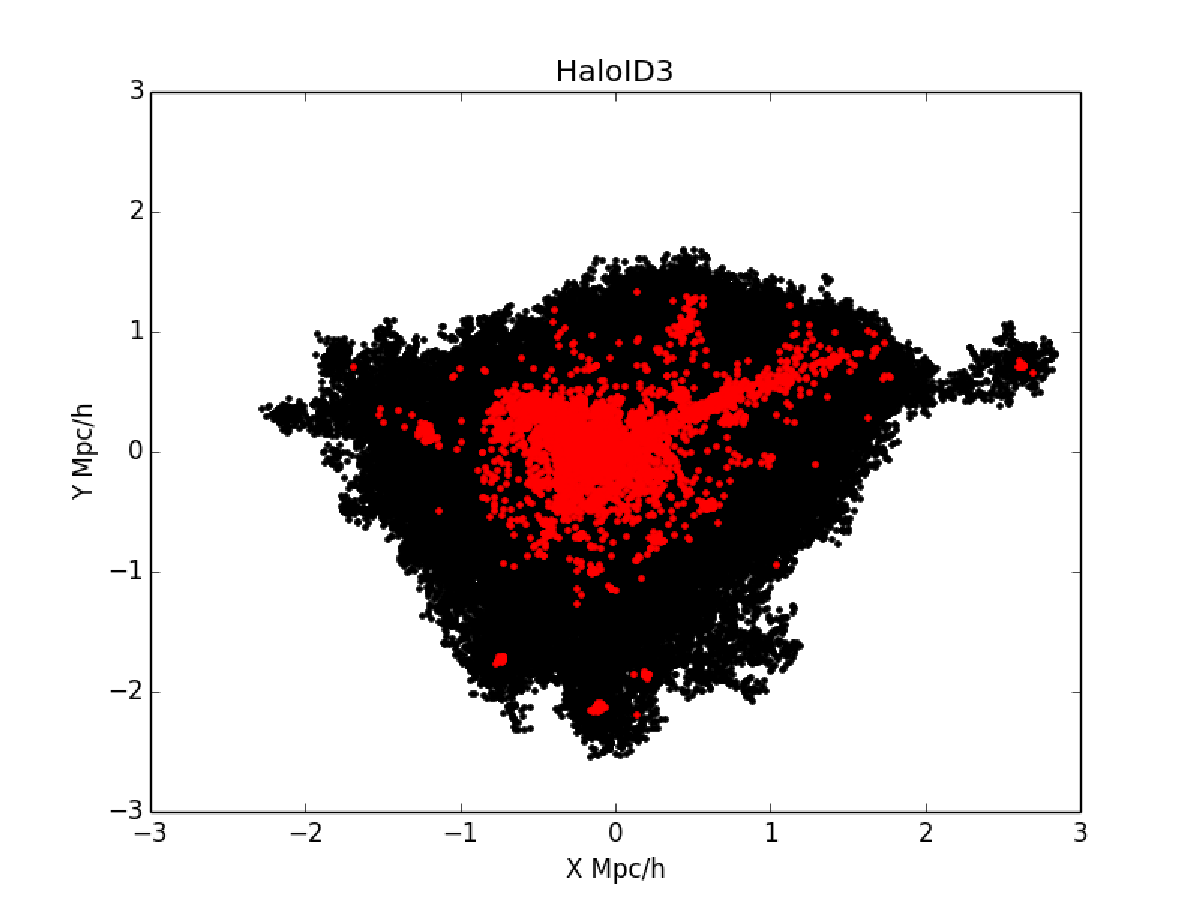}
\caption{An example of a halo from simulation. The black dots are the dark matter particles and the red ones are the stellar components. The radius is not regular due to the shape of the halo defined by $rockstar$ software.}\label{fig:halo}
\end{figure}

We also further test the possible contribution of faint satellites out of SDSS spectroscopic detection limit at r band model magnitude 17.77 around massive stellar mass bins, based on illustrisTNG300-3 \citep{nelson2018} low resolution hydro-simulation. IllustrisTNG-300-3 has 100 snapshots from z at 127,  with $302.6 \mpch$ box size, dark matter particle mass $3.8\times 10^9 \Msun$ and gas, stellar cell mass $7.0\times 10^8 \Msun$. We download group catalog from snapshot 91 at z=0.1 as well as processed offsets file to obtain the information of dark matter and gas, stellar particles for each halo and its subhalo. We select four samples based on halo mass (weak lensing mass$\pm error$) and stellar mass from Table \ref{tab:1}. The stellar particles inside 100kpc with respect to the centroids of the stacked dark matter particles, are considered to be from the central galaxies. This criteria is based on the ~50kpc offcenter effect \citep{luo_gg2} and the galaxy size 50kpc cited from \cite{chen2019}. The ratio between stellar particles outside this radius and the ones inside this radius is the rough estimation of the contribution of satellite galaxies in general. Fig.~\ref{fig:halo} is an example of halo from the simulation defined by $rockstar$ software \citep{behroozi2013ApJ}, the black dots are the dark matter particles and the red dots are the stellar particles, the boundary of the halo is not regular but roughly about the virial radius of a halo. We find ~10\% for the most massive stellar mass bin, and this dramatically decreases to 1.0\% for the second most massive stellar mass bin. This dramatic decrease may be due to the resolution of the suit of simulations we used here. However, we still can consider the 10\% as an upper limit for the satellite contribution. Further more,  in observational data, the secondary satellite is beyond 17.77 in r band, so in reality this is less than 10\%. And the contribution for the rest can be neglected. So the "unobserved" faint galaxies do not contribute significantly to the EG in our analysis. 

About 5.7\% galaxies (36, 759) in the sample are brighter than r band 17.77, but without spectroscopic 
redshift measurements due to fiber collision effect. According to \cite{zehavi2002}, roughly 60\% of the fiber-collision galaxies have a
redshift within 500 km s$^{-1}$. In \cite{yang2007}, they assign redshift of their nearest neighbours in the
group finding procedure. As a result, the single system does not have close companion with fiber collision galaxies, therefore our
results are not effected by fiber collision effect.

In a word, our results are robust against potential influence
from either fiber collision galaxies and faint galaxies with
r band magnitude fainter than 17.77.

\section{Summary and Discussion}
\label{summary}

We select isolated galaxy systems from SDSS DR7 group catalog \cite{yang2007}, with recent updated halo mass
estimation. This update doubles the number of lens galaxy at small stellar mass bins compared to the sample
used in \cite{chen2019}, which enables us to measure high SNR ESD for those samples(17.6 for blue galaxy sample to 28.1 for red galaxy sample). Further more, we 
split each stellar mass sample into blue and red to test the color dependence.

We model the ESD profile with NFW profiles, setting halo mass and concentration as free parameters based on two cosmologies, i.e. WMAP5 and PLANCK18.
The most significant difference is from the ESD between red and blue lens samples. The ESDs from the blue samples
 in the same stellar mass bin have lower amplitude than their red counterparts, indicating smaller halo mass.
 Because "apparent dark matter" ESD in EG framework remains the same as long as the stellar mass is the same.
 This can be clearly seen in stellar mass bin 2 and 3 where the stellar mass has only 0.014dex to 0.002dex
 difference, while the halo mass have up to ~5$\sigma$ difference. 
 
We also further test the validity of our selection of isolated systems using illustrisTNG300-3 \citep{nelson2018}, and we found that
 the contribution of possible satellite out of SDSS spectroscopic detection limit is ~10\% for the most massive stellar mass bin
 and 1\% for the second most massive stellar mass bin. This effect can be neglected for the rest of the samples.

In general, EG scenario of gravity failed to explain the color dependence of
the galaxy-galaxy lensing signal and we summarise as follows:
\begin{itemize}

\item The EG favors PLANCK18 cosmology with reduced $\chi^2_{reduced}=1.907$ to WMAP5 
$\chi^2_{reduced}=2.959$ with degrees of freedom of 15 for NFW and 25 for EG. The NFW model shows significant lower reduced $\chi^2$ value than those from EG already without red and blue dichotomy, which are 0.868(0.996) for WMAP5 cosmology and 0.949(1.453). 

\item The most significant difference is from the first three stellar mass bins
after the red and blue classification. For instance, in PLANCK18 cosmology the reduced $\chi^2$
is 0.718(0.885) for red lens sample and 0.659(0.626) for blue sample, and these values are increased to 1.792(1.762) and 2.730(2.931) respectively in EG model.

\item The halo mass discrepancy between abundance matching and NFW model fitting
is significant for the last two stellar mass bins, this is due to the combination
of selection effect and abundance matching method. 

\item Our results are consistent with \cite{zu2016} in that the halo mass of blue galaxies 
in the same stellar mass bins are smaller than that of red galaxies.
\end{itemize}

\acknowledgements
We are grateful to Prof. Y. F. Cai, K. Xu and H. Y. Wang from USTC and Prof. D. C. Dai from
YZU, for valuable comments. WL acknowledges the support from WPI Japan. All numerics
are operated on the computer clusters gfarm
at Kavli IPMU and cluster from Shanghai Astronomical Observatory. JZ is supported by IBS under the project code
IBS-R018-D1. XY is supported by the National Science Foundation of China (NSFC, grant Nos. 11890692, 11833005, 11621303) and the 111 project No. B20019. LL is supported by the National Science Foundation of China (NSFC, grant Nos. 11903067).

\bibliographystyle{apj}
\pagestyle{plain}
\bibliography{egtest}

\end{document}